\documentstyle[12pt,epsf]{article}
\newcommand{\EQ}{\begin{equation}}
\newcommand{\EN}{\end{equation}}

\newcommand{\EQA}{\begin{eqnarray}}
\newcommand{\ENA}{\end{eqnarray}}

\newcommand{\goto}{\rightarrow}
\newcommand{\be}{\beta}
\newcommand{\al}{\alpha}
\newcommand{\lam}{\lambda}
\def\acq{a_q^\dagger}
\def\aq{a_q}
\def\ra{\rangle}
\def\la{\langle}
\def\tr{{\rm Tr}}

\def\I{{\rm Im}\ }
\def\R{{\rm Re}\ }

\newcommand{\re}{\rm e}
\begin{document}
\oddsidemargin 5mm
\setcounter{page}{0}
\renewcommand{\thefootnote}{\fnsymbol{footnote}}
\newpage     
\setcounter{page}{0}
\begin{titlepage}
\begin{flushright}
DFF 289/9/97 \\
hep-th/9710071
\end{flushright}
\vspace{1.cm}
\begin{center}
{\Large {\bf Solving the Frustrated Spherical Model  \\
with $q$-Polynomials}}\\
\vspace{1.5cm}
{\bf Andrea Cappelli }\ \ \ {\em and} \ \ \ {\bf Filippo Colomo}\\
\vspace{0.8cm}
{\em I.N.F.N., Sezione di Firenze \\
and Dipartimento di Fisica, Universit\`a di Firenze, \\ 
Largo E. Fermi 2, 50125 Firenze, Italy}\\  
\end{center}
\vspace{2.cm}

\begin{abstract}
We analyse the Spherical Model with frustration induced by an external
gauge field. In infinite dimensions, this  has been recently
mapped onto a problem of $q$-deformed oscillators, whose real parameter $q$
measures the frustration.
We find the analytic solution of this model by suitably
representing  the $q$-oscillator algebra with 
$q$-Hermite polynomials.
We also present a related Matrix Model which possesses the same diagrammatic
expansion in the planar approximation. Its interaction potential
is oscillating at infinity with period $\log(q)$, and  may lead to
interesting metastability phenomena beyond the planar approximation. 
The  Spherical Model is similarly $q$-periodic, but does not exhibit
such phenomena: actually its low-temperature phase is not glassy and
depends smoothly on $q$.
\end{abstract}
\vfill
\vspace{5mm}
{\hfill September 1997}
\end{titlepage}


\section{Introduction}

In two recent papers  \cite{parisi1,parisi2}, Parisi {\em et al.} 
have introduced and analysed the Spherical and $XY$ Spin Models
with frustration, but in the absence of quenched disorder.
Their aim was to test the conjecture that the frustrated deterministic 
systems at low temperature behave as some suitably chosen 
spin-glass models with quenched disorder \cite{spinglass}.
They considered the frustrated models in the limit of large 
dimensionality $D$ of the lattice,
where the saddle-point approximation becomes exact.
In their analysis of these models, they showed that
the high temperature expansion can be nicely rewritten by using the
$q$-oscillators algebra  \cite{parisi1,parisi2}. Here
$q$ measures the frustration per plaquette and
varies continuously  between the fully-frustrated case 
($q=-1$, fermionic algebra)
and the ferromagnetic case ($q=1$, bosonic algebra).
Similar $q$-deformed algebraic relations have also appeared in
the Hofstadter problem of quantum particles hopping on a 
two-dimensional lattice in a magnetic field  \cite{wiegmann}.
This problem is closely related to the frustrated spin models,
which can be considered as simplified models of hopping in the large 
$D$ and classical limits. 
This relation provides another motivation for our analysis.
Finally, we note that the frustrated XY Model can also describe 
Josephson junctions  arrays in a magnetic field.

In this paper, we solve exactly the frustrated  Spherical Model 
in the large $D$ limit, both in the high and low temperature phases.
We use the ``coordinate'' representation of the $q$-oscillators, 
which is given by the $q$-Hermite polynomials \cite{maassen,gasper}.
We find that the spectrum of the lattice Laplacian is essentially given by a 
Jacobi theta function; thus, it is periodic along the imaginary axis,
with  period $\log(q)$. 
This $q$-periodicity does not affect 
the low-temperature phase of the Spherical Model, which is rather standard and 
non-glassy, the effect of frustration being quantitative only.
On the other hand, we show that the Spherical Model is associated to a 
Matrix Model  \cite{brezin,matmod}, which has the same diagrammatic 
expansion in the planar approximation. 
The potential of this Matrix Model is oscillating at infinity,
where it has infinitely many minima at approximate distance
$\log(q)$ (see Fig.(2)). 
Once the corrections to the planar limit are considered,
its states become metastable by tunnelling to these
minima: therefore, we argue that this model could have a band spectrum and 
possibly behave as a spin glass. 
Similar phenomena could also occur in the frustrated $XY$ Model,
whose analysis is however left to future investigations.

\bigskip

The models we consider are described by the Hamiltonian
\EQ
H_0=-{1\over\sqrt{2D}}\ \sum_{\la j k \ra} \phi^\dagger_j U_{jk}\phi_k \  
+ \  {\rm h.c.} \ .
\label{ho}
\EN
The complex field $\phi_j\in {\bf C} $ is
defined on the sites (labelled by $j$) of a $D$-dimensional hyper-cubic lattice,
and there are nearest-neighbour interactions $U_{ij}$.
Three different models can be  obtained by constraining the field as
follows:
\EQA
\be H_G&=&\be H_0 +\sum_{j=1}^N|\phi_j|^2 \ ,\label{hgauss}\\
\be H_S&=&\be H_0 +\mu\left(\sum_{j=1}^N|\phi_j|^2 \  -N\right) \ ,
\label{hsfer}\\
\be H_{XY}&=&\be H_0 + \sum_{j=1}^N\ \mu_j\left(|\phi_j|^2 \  -1\right) \ .
\label{hxy}
\ENA
The first is the Gaussian Model, which only exists in the 
high-tempe\-rature phase, where the mass term  dominates the 
kinetic term (\ref{ho}). The second is the Spherical Model  
\cite{berlin}, which uses the 
Lagrange multiplier $\mu$ to
enforce the condition $\sum |\phi_j|^2=N$, 
$N$ being the total number of sites. 
This Lagrange multiplier is promoted to a field $\mu_j$ in 
the $XY$ Model (\ref{hxy}), whose variables are constrained to  live on
the unit circle $|\phi_j|=1$, $\ \forall j$.

The couplings $U_{jk}$ are complex numbers of modulus one and satisfy the
relation $U_{jk}=U^*_{kj}$; they are the link variables of an Abelian 
lattice gauge field, without kinetic term, which produces a static external 
magnetic field. This is chosen in such a way as to give 
the same magnetic flux $\pm B$ for
any plaquette of the lattice (the product of the four $U$'s 
around the plaquette is  $e^{\pm i B}$). Therefore, 
the magnetic field should have the same
projection on all the axes of the lattice, modulo the sign. 
In order to avoid the choice of a preferred direction in the lattice,
these signs are chosen randomly\footnote{
This is a small amount of randomness, of order
$D(D-1)/2$, to be compared to the $\sim L^D$ randomness present 
in  systems with quenched disorder.}  \cite{parisi1}.

The ferromagnetic spin interaction is obtained for $B=0$, i.e.
$ U_{jk} =1$. Non-vanishing values of $B$ induce a frustration
around each plaquette, which is maximal for $B=\pi$,
the fully-frustrated case. Of course all the intermediate, partially
frustrated cases, with $0<B<\pi$, are  also interesting.
As explained in Ref.  \cite{parisi1}, the $B=0$ theory should be treated 
with care, because the large $D$ limit is different in this case, 
and the normalisation factor $1/\sqrt{2D}$ in (\ref{ho}) should be 
replaced by  $1/(2D)$.

In order to study the previous models, the main difficulty consists
in finding the spectrum of the lattice Laplacian in the presence of  the
magnetic field, which is defined as:
\EQ
(\Delta f)_j=\sum_{k=1}^{2D} \  U_{jk} \  f_k \ .
\EN
Similarly to
the Hofstadter problem  \cite{wiegmann}, there is a competition between the 
periodicity due to the lattice and those induced by commensurate magnetic 
fields of the form $B=2\pi r/s$, with $r,s$ integers. Thus, we could
expect a complex band structure in the spectrum.
However, there are simplifications due to the large $D$ limit.
The authors of Refs.  \cite{parisi1,parisi2} have approached the
problem via the high-temperature expansion.
In the case of the free energy of the Gaussian Model 
(eq. (\ref{hgauss})), this is given by a sum over all closed loops,
as follows:
\EQ
\be F=\sum_{n=0,\ {\rm even}}^\infty \ \ \ {1\over n}\  
\left(\frac{\be}{\sqrt{2D}}\right)^n \  {\cal N}(n)\  \la W ({\cal C}) \ra_n
=\sum_{k=0}^\infty\ {\be^{2k}\over 2k}\ G_k \ .
\label{htexp}
\EN
In this expression, the loops ${\cal C}$ are arranged according to their 
length $n=2k$, and their number is ${\cal N}(n)$.
For each loop, the magnetic field yields a weight, which is given by
the Wilson loop $W({\cal C})$, the path-ordered product
of the couplings $U$ along the loop.
The brackets $\la \ \ra_n $ represent the average of this weight
over all the ${\cal N}(n)$ circuits of length $n$.
In the second expression of eq. (\ref{htexp}), we introduce the
notation $G_k$ for the product of the multiplicity
and the Wilson loop average. 

Each loop encloses a number of plaquettes and receives a weight
proportional to $\exp{(i B A)}$, where $A$ is the sum of plaquettes
with signs depending on the orientations  \cite{selfcit}. 
Due to the average over orientations and loops, the quantity 
$G_k$ is a polynomial in the variable
\EQ
q= \cos B \ ,
\label{qdef}
\EN
of order ${k(k-1)/2}$, which is given by the maximal area 
enclosed by the loop.
Parisi  \cite{parisi1} enumerated these 
diagrams in the large $D$ limit.
First, he showed that the same counting is given by 
the Feynman diagrams with $2k$ external points, which are joined pairwise 
by lines (propagators) intersecting $I$ times, and have assigned the
weight $q^I$. 
These diagrams also occur in the topological (large $N$) expansion of
Matrix Models \cite{brezin}, where  the planar limit corresponds to no 
intersections, i.e. to the $q=0$ case. 
Secondly, Parisi found a recursion
relation for the coefficients of the polynomial $G_k(q)$
-- a sort of Wick theorem -- which
can be nicely  expressed by the algebra of the $q$-oscillators  
$a_q, a_q^\dagger$:
\EQ
\aq\acq- q \acq\aq=1 \ .
\label{qalg}
\EN
These operators \cite{kulish} act on the Hilbert space spanned by the vectors: 
$|m\ra$, $m=0,\ 1,\ ... $, as follows,
\EQA
\acq |m\ra  &=& \sqrt{[m+1]_q}\ |m+1\ra \ ,\nonumber\\
\aq |m\ra  &=& \sqrt{[m]_q}\ \quad |m-1\ra \ ,\qquad \aq |0\ra= 0 \ ,
\ENA
where
\EQ
[m]_q=\frac{\  1-q^m}{\  1-q\ } \ .
\EN
Using the recursion relation, the weighted multiplicities of the diagrams 
of eq. (\ref{htexp}) were neatly written as an expectation value over 
the ground state of 
the $q$-oscillators  \cite{parisi1,parisi2}:

\EQ
G_k(q) \ =\ \la 0|(\acq+\aq)^{2k}|0\ra \ .
\label{qvev}
\EN

\section{Spectrum of the Laplacian and High-Temperature Expansion}

We shall now make  expression (\ref{qvev}) more explicit by introducing
the coordinate representation for the $q$-oscillators.
First note that in the Gaussian Model (\ref{hgauss})
the quantity $G_k$ is nothing else then the trace of $(2k)$-th power of
the frustrated Laplacian, such that we can write the general relation
\EQ
 \tr \ \left[ f\left(\Delta\right)\right] \equiv \la 0 |  f(x_q) |0\ra\ ,
 \qquad x_q\equiv \acq + \aq \ .
 \label{qveve}
\EN

The $x_q$ coordinate representation, $x_q|x\ra = x|x\ra$, 
is given by the so-called continuous $q$-Hermite polynomials  \cite{gasper,szego}.
These are defined by:
\EQ
H_n(x) =  \la x | n \ra\ {\cal C}_n\ ,
\qquad {\cal C}_n = \left(\left[n\right]_q!\right)^{1/2}\ {\cal C}_0\ ,
\EN
where the normalization constant ${\cal C}_0$ is fixed by
$H_0(x) =1 $ and the $q$-factorial is 
\EQ
\left[n\right]_q! =[n]_q \ [n-1]_q \dots [1]_q\ , \qquad
[1]_q=[0]_q=1 \ .
\EN
These polynomials satisfy, of course, a three-term recursion relation
in the index $n$: 
\EQ
x H_n(x) = H_{n+1}(x) +[n]_q\ H_{n-1}(x)\ , \qquad n\ge 1 \ .
\EN
Moreover, they obey a $q$-difference equation \cite{wiegmann}\footnote{
This $q$-periodicity will be better discussed later.}
in their coordinate $x$, which ranges over the interval 
$x\in\left[-2/\sqrt{1-q},2/\sqrt{1-q}\right]$.
A convenient parametrisation is
\EQ
x=\frac{2}{\sqrt{1-q}}\cos\theta \  ,\qquad \theta\in [0,\pi ].
\label{variasm}
\EN
More properties of these $q$-Hermite polynomials can be found
in  Ref.  \cite{gasper}, where they are defined as 
${\cal H}_n(\cos\theta)=(1-q)^{n/2} H_n (x)$.
The most important property for us is the orthogonalizing measure
$\nu_q(x)\ \ $  \cite{maassen,gasper,szego}:
\EQ
\int_{-2/\sqrt{1-q}}^{2/\sqrt{1-q}}\ \nu_q(x)\ dx\ 
H_n(x)\ H_m(x) \ = \  \delta_{n,m}\ [n]_q! \ ,
\EN
\EQA
\nu_q(x) &=&\frac{\sqrt{1-q}}{2\pi}\  q^{-1/8}\  
\Theta_1\left({\theta\over\pi},q\right)\  \nonumber\\
&=&\frac{\sqrt{1-q}}{\pi}\ \sum_{n=0}^{\infty}\ (-1)^n q^{n(n+1)/2}
\ \sin[(2n+1)\theta] \nonumber\\
&=&\frac{\sqrt{1-q}}{\pi}\ \sin\theta\ 
\prod_{n=1}^{\infty}\ \left(1-q^n \right)\ 
\left(1-q^n \re^{2i\theta} \right)\ \left(1-q^n \re^{-2i\theta} \right) \ ,
\label{qmeasure}
\ENA
where $\Theta_1(z,q)$ is the first Jacobi theta function.
 
Using these results, we can rewrite the general ground-state
expectation value in eq. (\ref{qveve}) as follows:
\EQ
\tr \left[ f\left(\Delta\right)\right] \equiv 
\int\ dx\ \nu_q(x)\ f(x) \ .
\EN
Therefore, we have shown that the Laplacian is diagonal in the coordinate 
representation of the $q$-oscillators: it has a continuous spectrum
over the interval $[-2/{\sqrt{1-q}}, 2/{\sqrt{1-q}}]$,
with eigenvalue density given by $\nu_q$ in eq. (\ref{qmeasure}).

Let us discuss this spectrum in some interesting limits.
For $q=0$, we find $\nu_0(x)=1/\pi\ \sqrt{1-x^2/4}$, which is 
the Wigner semi-circle law for the Gaussian Hermitean Matrix 
Model  \cite{brezin}.
This results confirms the previous correspondence between
the high-temperature expansion of the $q=0$ Gaussian Model and the  
planar Feynman diagrams without interaction vertices of the Matrix Model. 
Actually, this correspondence can be extended to all values of $q$
(see Section 4).
Note also that the $q=0$ frustrated
model is diagrammatically equivalent to the gauge spin glass where
the coupling $U_{jk}$ are random quenched variables  \cite{parisi1}.
Next, for $q=-1$, the measure becomes a
representation for $(\delta(x-1)+\delta(x+1))/2$, and we recover
 the two states of the fermionic algebra \cite{maassen}.
The limit $q\goto 1$  is singular, owing to the 
$D\to\infty$ peculiarities said before; nevertheless,  
$\nu_1(x)$ becomes the Gaussian distribution, after multiplicative 
renormalization  \cite{maassen}.

Let us now check that our expression for the Laplacian reproduces 
the high temperature expansion computed in
Refs.  \cite{parisi1,parisi2}.
The free energy of the Gaussian Model (\ref{hgauss}) can be written in
terms of the Laplacian as $\be F=-\log Z= \tr [\ln (1-\be\Delta)]$.
Therefore, the internal energy $U(\be)$ is:
\EQ
1-\be U(\be)=R(\be)\ ,\qquad 
R(z)\equiv \tr\left[\frac{1}{1-z\ \Delta}\right] \ .
\label{resolvent}
\EN
In the last equation, we introduced
the resolvent $R(z)$, which will play a major role in the subsequent
study of the Spherical Model.
$R(z)$ is originally well-defined for real $z$ in the interval 
$|z|<z_c=(\sqrt{1-q})/2$ and is then  analytically extended to the 
whole complex plane of $z$.
By using the results of the previous section, we find:
\EQA
R(z)&=&\int_{-2/\sqrt{1-q}}^{2/\sqrt{1-q}}
\  dx \ \frac{\nu_q(x)}{1-z\ x} \nonumber\\
&=&{\sqrt{1-q}\over z}\ \sum_{n=0}^{\infty} \ (-1)^n\ q^{n(n+1)/2} \ 
\left[K\left(\frac{2z}{\sqrt{1-q}}\right)\right]^{2n+1} \ ,
\label{erre}
\ENA
where we introduced
\EQ
K(\al) = \frac{1 - \sqrt{1-\al^2}}{\al}\ .
\label{gi}
\EN
The singularities of $R(z)$ in the complex plane 
are completely determined by those of the simpler function 
$K(2z/\sqrt{1-q})$. 
Actually, the series is very well convergent for $|q|<1$ as any Jacobi theta
function. Moreover, for $|q|=1$, it is a geometric series, 
which is still convergent because $|K| < 1$ for $|z|<z_c$.

The high-temperature expansion of the internal energy of the Gaussian
Model can be obtained by expanding $R(\be)$ in series;
there only appear even powers of $\be$ with $q$-dependent coefficients.
We have computed the series to order $O(\be^{18})$ 
with Mathematica  \cite{math}, and obtained the polynomials
$G_k(q)$ , $k=1,\dots,9$ of eq. (\ref{htexp}).
We have verified that they indeed match the results of 
Refs.  \cite{parisi1,parisi2}, which were found by direct enumeration of the
graphs on a computer.

For $q=0$, we have $U(\be)=\left(\be - K(2\be) \right)/\be^2$ and the 
high-temperature expansion is singular at $\be_c=1/2$, which is 
the critical
temperature of the Gaussian Model. The specific heat diverges 
as $(\be_c-\be)^{-1/2}$ at the transition, $i.e.$ the 
value of critical exponent $\alpha$ is $1/2$. 
This result still holds for generic $-1<q<1$ as a consequence of the
relation between the singularities of $R(z)$ and $K(z)$.

\section{Solution of the Spherical Model}

We are now ready to solve the Spherical Model. Our approach
essentially follows the original Berlin-Kac solution of the ferromagnetic 
model in 
$D=1,2,3$  \cite{berlin}. Indeed we shall see that the mechanism of the phase
transition is the same, only the form of the Laplacian is different.
The partition function is
\EQA 
Z&=&\int_{\mu_0-i\infty}^{\mu_0+i\infty}d\mu\ 
\int{\cal D}\phi \ \exp\left[ \be \sum\ \phi^\dagger\Delta\phi
                        -\mu\left(\sum |\phi|^2-N\right)\right] \nonumber\\
&=&\int_{\mu_0-i\infty}^{\mu_0+i\infty} d\mu\ \
\exp\left[-\tr[\ln(\mu-\be\Delta)]+\mu N \right] \ .
\label{partsfer}
\ENA
The integration path over $\mu$ is along a straight line which runs 
in the complex plane parallel to the imaginary axis; a positive real
part has been added in order to make the integration over $\phi$
convergent.
The trace in the exponent is proportional to $N$ ($\Delta$ is an 
$N$-dimensional operator): in the large $N$ limit, we can apply the 
saddle point method to evaluate the integral over $\mu$.
The saddle point equation is:
\EQ
\tr\left[\frac{1}{\mu-\be\Delta} \right]=1\ ,
\label{saddl}
\EN
and must be solved for real, positive values of $\mu$. Introducing the
variable $z=\be/\mu$, this equation can be
written in terms of the resolvent (\ref{resolvent}), as follows:
\EQ
\frac{\be}{z} = R(z) \ ,\qquad z\equiv \frac{\be}{\mu} \ .
\label{saddle}
\EN

In the case of $q=0$, this equation can be easily solved:
\EQ
\frac{1}{2z^2}(1-\sqrt{1-4z^2})=\frac{\be}{z} \qquad \longrightarrow \qquad
z(\be)=\frac{\be}{1+\be^2} \ ,
\label{spq0}
\EN
which corresponds to $\mu=1+\be^2$.
This solution is valid for $z<z_c=1/2$, namely
it describes the high-temperature phase $\be<\be_c=1$. 
Indeed, let us follow the saddle point in the $z$ plane,
starting from $\be=0$, where $z=0$ too. Upon increasing 
$\be$, $z$ moves toward the critical value $z_c=1/2$;
for $\be >1$, $z$ cannot increase anymore because it finds 
the square-root branch cut of $R(z)$, 
and therefore ``it sticks to the singularity''  \cite{berlin}.
Hence, for $\be>1$, the saddle point equation has no longer an
acceptable solution; nevertheless, the leading contribution to the 
$\mu$ integral in (\ref{partsfer}) still comes from the neighbourhood of 
$z=1/2$. 
Therefore, the low-temperature solution is $z(\be)=1/2$ independent of $\be$,
$i.e.$ $\mu=2\be$.
The free and internal energies are therefore given by
\EQA
\be F&=&\tr[\ln(\mu-\be\Delta)]-\mu \ ,\\
U(\be)&=&(\partial_{\be}+(\partial_{\be}\mu)\partial_{\mu})(\be F) 
\nonumber\\
&=&\frac{1}{\be}\left(1-\tr\left[\frac{1}{1-z\Delta}\right]\right)+
\left(\frac{1}{\mu}\tr\left[\frac{1}{1-z\Delta}\right]-1\right)
{\partial\mu \over\partial\be}\ .
\ENA
The last term in this equation vanishes in  the high-temperature phase, 
because the saddle point condition is satisfied; however, it should
be included for the low temperature regime. 
As a matter of fact, the result can be written in both phases as follows:
\EQ
U(\be)=\frac{1}{\be}-\frac{1}{z(\be)} \ .
\label{usfer}
\EN
More explicitly,
\EQ
U(T)=\left\{
\begin{array}{lll}
-\frac{1}{T}  & & T>1 \\
T-2           & & T<1
\end{array} \right. 
\qquad \left( q=0 \right ) \ .
\EN
The internal energy is therefore continuous at the transition,
together with its first derivative. But the specific heat presents  a 
discontinuity in its first derivative. These results are qualitatively
very similar to those of the Berlin-Kac analysis  \cite{berlin};
ne\-ver\-the\-less, their model is quite different, in the sense that
frustration is there completely absent.

The solution for any $-1< q < 1 $ is a straightforward generalization
of the $q=0$ case. The discussion of the saddle-point equation (\ref{saddle}) 
is analogous, because the singularities of $R(z)$ 
are still given by a square-root branch cut. Although we cannot find
an explicit expression for $z(\be)$, we can follow its behaviour:
starting from low $\be$, where $z\sim \be$, the 
saddle point moves towards higher values of $z$ until it hits
the cut of $R(z)$ at $z=z_c=\left(\sqrt{1-q}\right)/2$.
In the low-temperature phase, the value of $z$ sticks to $z_c$
and thus $\mu=\be/z_c$ for $\be>\be_c$. 
The critical temperature is given by
\EQ
\be_c=z_c R(z_c)=\sqrt{1-q}\ \sum_{n=0}^{\infty} (-1)^n q^{n(n+1)/2} \ .
\EN

The internal energy is still given by the general formula (\ref{usfer}), 
which now reads
\EQ
U(T)=\left\{
\begin{array}{lll}
T - \frac{2}{\sqrt{1-q}} & & T<T_c \\
- \frac{1}{T} -q\frac{1}{T^3} -\left(q^3+3q^2\right)\frac{1}{T^5}
+ O\left( \frac{1}{T^7} \right)          & & T \gg T_c
\end{array} \right. 
\qquad \left( -1< q<1 \right ) \ .
\EN
\begin{figure}
\epsfxsize=15cm \epsfbox{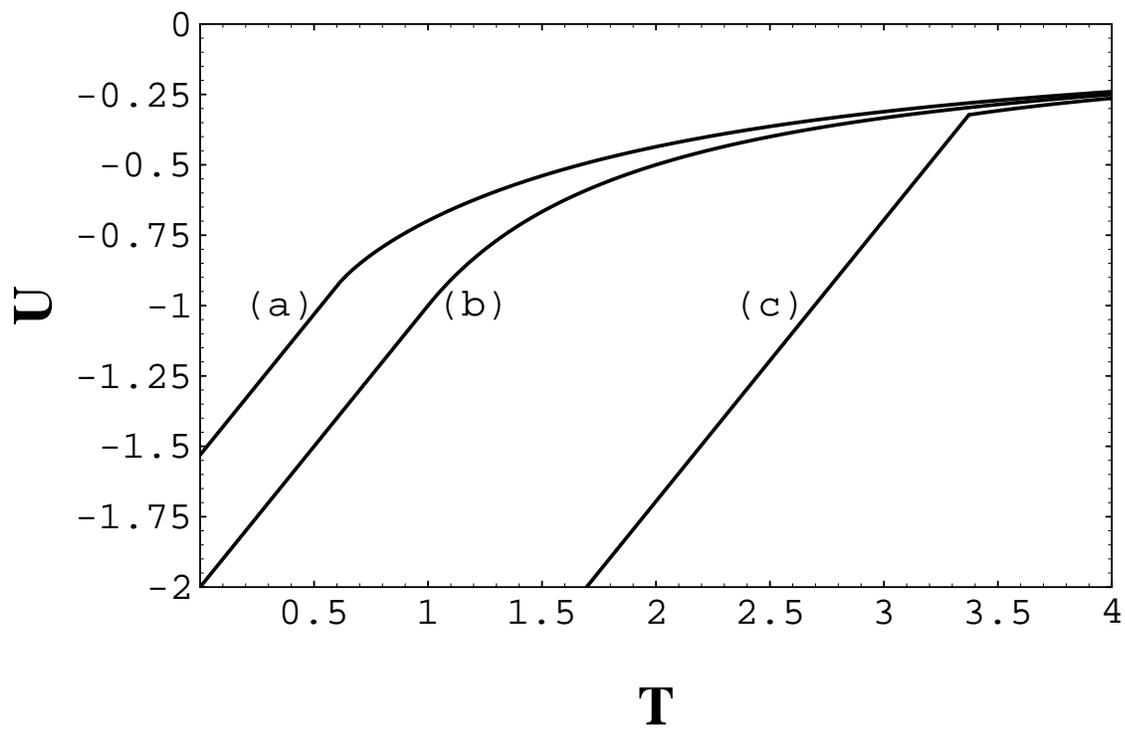}
\caption{Internal energy of the Spherical Model  versus 
temperature for the three values: $(a)$ $q=-0.707$; $(b)$ $q= 0$;
$(c)$ $q=0.707$.} 
\label{fig1}
\end{figure}
A plot of the internal energy is shown in Fig.({\ref{fig1})
for the three values $q=(-1/\sqrt{2}, \ 0, \ 1/\sqrt{2})$. 
This can be easily obtained without solving the saddle-point equation,
by using $z\in [0,z_c]$ as a parameter
for both $U(\be)$  in eq. (\ref{usfer}) and $\be$ in eq. (\ref{saddle}).
Note that for $q\neq 0$, the first derivative of the internal energy is not
continuous at the transition. The critical temperature goes to
infinity for $q\to 1$ and to zero for $q\to -1$, respectively.
In these limiting cases, we can sum the series in $R(z)$ (\ref{erre}) 
and obtain the explicit form of $z(\be)$ and the internal energy.
For $q=1$, we find the meaningless results $z=\be$ and $\ U(T)=0$ 
in the high-temperature phase, which collapses to a point ($T_c=\infty$): 
this case has a pathologic $D\to\infty$ limit, 
as discussed in Ref.  \cite{parisi1}, and a sensible theory should include
$1/D$ corrections.

The limit $q=-1$ is well defined and 
corresponds to the fully frustrated model at $D=\infty$\footnote{Note
that the limits $D\goto\infty$ and $q\goto -1$ do not commute,
as discussed in Ref.  \cite{parisi1}.}. 
Here, the high-temperature phase extends down to $T=T_c=0$, i.e. there is
no transition. Actually, we find that the resummed $R(z)$ no longer
has the square-root branch cut of $K(z)$ and that the mapping
$\be=\be(z)$ is invertible on the whole positive $\be$ axis:
\EQA
\be &=&  \frac{z}{1-z^2} \ , \qquad\qquad \qquad \qquad 
\left( q=-1\right) \nonumber\\
U &=& -z =  \frac{1}{2\be}\left(1- \sqrt{1+4\be^2} \right) \ .
\ENA
Let us remark that the absence of a phase transition 
has also been observed in the fully-frustrated
long-range $XY$ model  \cite{fermi}.

In conclusion, the qualitative behaviour of the (infinite-dimensional)
frustrated Spherical Model in the low-temperature phase is rather smooth and 
standard: indeed, this model does not present the potential features of:
(i) glassy behaviour (many ground states), which 
is usually found in  systems with quenched disorder; 
(ii) commensurability in the spectrum for $B=2\pi r/s$,
which was observed in the $D=2$ hopping model in a magnetic field.
Regarding point (i), we would like to remark  that the corresponding
model with quenched disorder, $i.e.$ the Spherical Gauge Glass,
does not exhibit a glassy phase either. Actually, this model corresponds
diagrammatically to the $q=0$ frustrated model. In intuitive terms, the 
spin variables are only loosely constrained by the spherical condition,
$\sum_{i} |\phi_i|^2 = N$, and thus can globally adapt themselves to any 
complex coupling configuration. As far as point (ii) is concerned, the
commensurability effects are probably washed out  by the $D\goto\infty$ limit,
which oversimplifies the geometry of hopping on the lattice.

On the other hand, we expect a glassy phase in the frustrated XY Model
\cite{parisi2}. 
In this model, the Lagrange multiplier $\mu$ becomes a local 
field $\mu_j$ and the saddle-point equation (\ref{saddl}) is functional. 
Nothing changes in the high-temperature phase,
where the solution $\mu_j=\mu=$ const. is correct.
However, the solution(s) in the low-temperature phase can be rather different
from the one of the Spherical Model, and will not be discussed here.
Rather, we shall approach this problem from a different
perspective, by establishing a relation with the well developed 
subject of Matrix Models.

\section{Analogy with the Matrix Models}

Let us first recall the solution of the Hermitean Matrix models 
in the the large $N$ approximation, which corresponds to the 
planar Feynman diagrams  \cite{brezin}.
These models are characterized by the $(N\times N)$  matrix variable
$M=M^\dagger$ and by the Hamiltonian
$\be H=\tr V(M)= \tr \left( M^2 /2 + \cdots \right)$.
After diagonalization of the matrix, one is left with the partition
function over the eigenvalues $\lam_i$, $i=1,\dots , N$ :
\EQ
Z_M =\int \prod_{i=1}^N \ d\lam_i\ {\re}^{-\sum_i V\left(\lam_i \right)}\ 
\prod_{i< j} \left(\lam_i -\lam_j \right)^2\ .
\EN
This can be thought of as being the statistical mechanics of $N$
charges with coordinates $\lam_i$ in one dimension,
which repel each other logarithmically and are kept together
by the external potential $V(\lam)$. 
This one-dimensional gas of charges usually has a unique phase.
In the large $N$ limit, 
the partition sum is dominated by the contribution of the saddle point,
which corresponds to the equilibrium configuration of the charges,
neglecting fluctuations.
Moreover, the charges become a continuum with density $\nu(\lambda)$, 
which is normalised to one by a convenient rescaling  \cite{brezin}.
The saddle point equation is:
\EQ
\frac{1}{2}\ V^\prime (\lambda) =
P \int _{-a}^{a} dx\ {\nu(x) \over \lam -x} \ ,\qquad
\lam \in (-a,a) \ , \qquad \int _{-a}^{a} dx\ \nu(x) = 1 \ ,
\label{saddlemat}
\EN
where $P$ stands for the principal value of the integral.
This is an equation for the unknown $\nu(x)$, as a function of
the given potential $V(\lam)$. 
Following Ref.  \cite{brezin}, we introduce the function
\EQ
F(\lam) =\int_{-a}^a \ dx\ {\nu(x) \over \lam -x} \ ,
\label{effe}
\EN
with $\lam$ taking values in the complex plane. This 
function is analytic outside 
the segment of the real axis $(-a,a)$ corresponding to 
the spectrum; moreover, it goes to zero
at infinity as $1/\lam$, due to the normalisation of $\nu$.
For $\lam$ inside the spectrum, we have
\EQ
\R F(\lam) =\frac{1}{2} V^\prime (\lam)\ ,\qquad
\I F(\lam) = -\pi \nu(\lam) \ ,\qquad \lam\in (-a,a) \ .
\label{reimeffe}
\EN
The saddle point equation is thus equivalent to these relations for 
the function $F(\lam)$: they can be usually solved by analyticity arguments, 
and determine the density $\nu(\lambda)$.

Here we would like to remark that
these formulas are rather similar to those encountered 
in the Spherical model. Actually, we can identify the two saddle-point
equations (\ref{saddle}) and (\ref{saddlemat}) as follows:
\EQ
F(\lam)\equiv \frac{1}{\lam}\  R\left( \frac{1}{\lam} \right) \ ,
\qquad z\equiv \frac{1}{\lam} \ .
\label{relation}
\EN
More precisely, the saddle point equation for the Spherical Model is
discussed for $\lambda$ outside the spectrum, $\nu(\lam)$ is given and
$\be=\be(z)$ is the unknown.
On the other hand, in the Matrix Model $\lam$ is inside the
spectrum, $V(\lam)$ is given and $\nu(\lam)$ is the unknown.

\begin{figure}
\epsfxsize=15cm \epsfbox{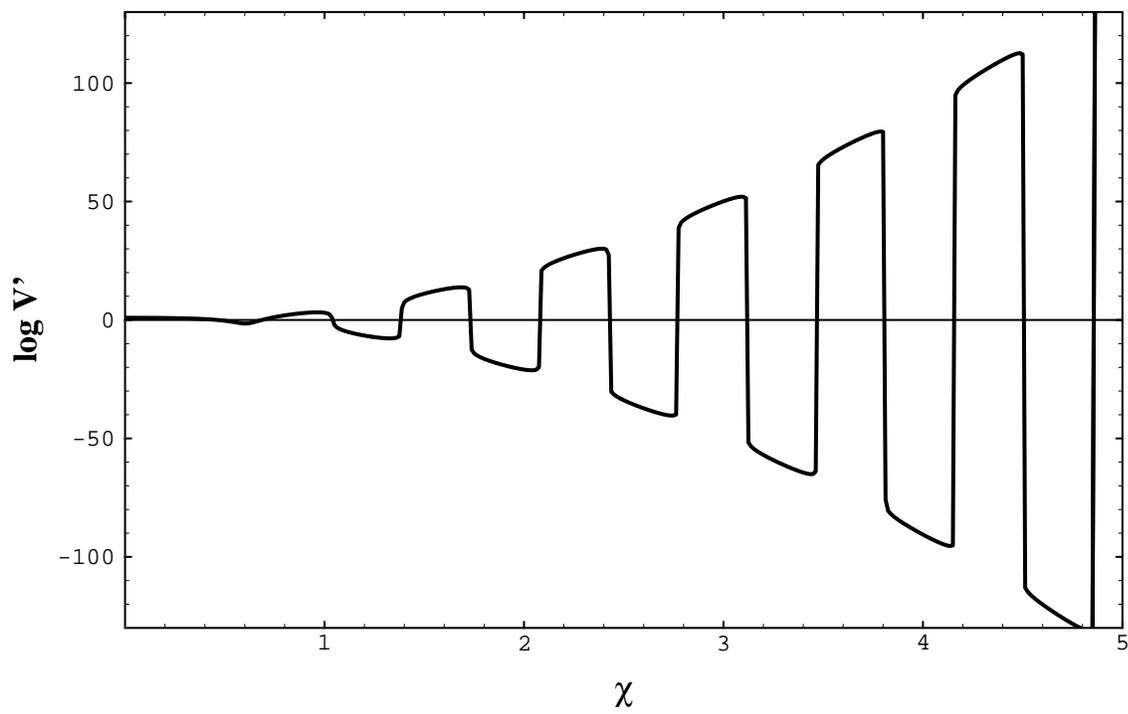}
\caption{The first derivative of the Matrix Model potential 
in eq. (39) as a function of $\chi$, for $q=-0.707$.}
\label{fig2}
\end{figure}

We can now define the Matrix Model corresponding to
the frustrated Spherical Model, as the one which possesses   
the same eigenvalue density $\nu_q(\lambda)$, eq.  (\ref{qmeasure}), 
in the planar limit.
By using  eqs. (\ref{effe}) and (\ref{reimeffe}), we can determine 
its potential:
\EQ
V^\prime \left( \lam \right) = 2 \R F(\lam) = 
2 \sqrt{1-q} \ \sum_{n=0}^\infty (-1)^n q^{n(n+1)/2} 
\cosh \left[(2n+1)\chi\right] \ ,
\label{vprime}\EN
where we introduced the convenient parametrisation 
\EQ
\lam = \frac{2}{\sqrt{1-q}} \cosh\chi \ ,
\label{variamat}
\EN
for $\lam$ outside the cut $(-2/\sqrt{1-q},2/\sqrt{1-q})$.
The Gaussian Matrix Model with $V(\lam)=\lam^2/2$ is indeed
recovered for $q=0$.

The relation (\ref{relation}) between the Spherical and Matrix Models can also
be understood at the level of diagrammatic expansions.
In the former model, $R(z)$ generates the high-temperature expansion,
whose diagrams were cast into the form of $n$ lines joining $2n$ points
with weight $q$ for each intersections  \cite{parisi1}.
In the latter model, $F(1/z)/z$ is the generating function for the observables
$\la \tr M^{2n} \ra= \la \sum_i\lam_i^{2n} \ra$  \cite{brezin}.
Their diagrams also have $2n$ external points: in the Gaussian Model ($q=0)$,
they are joined by propagator lines with no intersections, due to
the planar limit. For $q\neq 0$, the Spherical Model diagrams
have non-planar intersections, which are reproduced by interaction
vertices in the planar Matrix Model diagrams; actually, the potential 
$V(\lam)$ contains interactions of any order.

The Spherical Model in the high-temperature phase and the Matrix Model in 
its planar limit have corresponding saddle point equations but
rather different free energies, so they are not quite the same 
physical problem. 
The analogy with Matrix Models is, nevertheless, interesting because it
could provide some useful technology  \cite{matmod} for solving the 
$XY$ Model. 
Furthermore, the Matrix Model itself could exhibit 
interesting physics beyond the planar approximation:
in the following, we shall put forward some educated guesses
which are based on the well-known physical picture of the gas of charges.

The properties of the Matrix Model potential $V(\lam)$ can be found by
analysing equation (\ref{vprime}). A plot of $V^\prime(\chi)$ is shown
in Fig. (\ref{fig2}), for $q=-1/\sqrt{2}$: 
away from the origin, an oscillating behaviour sets in 
with period\footnote{
The period is $|\log q|/2$ for $0<q<1$.}
$\chi \to\chi -(\log q)$, and amplitude growing to infinity.
This {\it q-periodicity} of the potential can be found by inspection 
of (\ref{vprime}), and reads:
\EQ
V^\prime\left( q^{-1/2}\re^\chi\right) = 
2 \left(\frac{1-q}{q}\right)^{1/2} \re^{\chi} -
q^{-1/2}\re^{2\chi}\ V^\prime\left(\re^\chi\right) \ .
\EN
This is not an exact periodicity, due to the presence of
the additive term. However, this term becomes negligible for $\chi$ far away 
from the origin, i.e. $\lam\gg 1$, because $V^\prime$ grows
more than exponentially (the amplitude of fluctuations is of order
$O\left(\exp(2\chi^2/|\log q|)\right)$).
The $q$-periodicity of the potential corresponds to a true
periodicity of the eigenvalue density 
$\nu_q(\theta)\propto\Theta_1(\theta/\pi,q)$, 
because these two quantities are the real and imaginary parts of
the function $F$ in (\ref{effe}), respectively (their variables 
$(x,\theta)$ in (\ref{variasm}) and $(\lam, \chi)$ in (\ref{variamat}) 
are also related by analytic continuation).
Actually, the eigenvalue density is periodic in the direction of
the imaginary $\theta$-axis: 
$\nu_q(\theta - (i/2)\log q)\sim\nu_q(\theta)$, for $0<q<1$, 
and $\nu_q(\theta +\pi/2 -(i/2)\log q)\sim\nu_q(\theta)$, for $-1<q<0$.

This periodicity can be interpreted as the 
potentiality for metastable states, which, however, are not realised in 
the Spherical Model, owing to its simplified dynamics.
Indeed, its saddle point equation involves the function $F(\lam)\ $,
which is also $q$-periodic, as  its real part $V^\prime$; 
however, $F(\lambda)$ never grows sufficiently high to develop the 
oscillating behaviour
and, in fact, goes monotonically to zero at infinity.

On the other hand, the states might become metastable in the Matrix Model
beyond  the planar approximation. In this approximation, the charges form
an equilibrium configuration determined by the minimum of
$V(\lambda)$ at the origin, and the tunnelling of charges to lower nearby
minima is suppressed \cite{matmod}: 
thus, the $q$-periodicity of the potential is
not felt. Beyond this approximation, tunnelling switches on and the states in
the spectrum can become metastable. 
Clearly, a detailed analysis is necessary to understand the effect
of tunnelling into a $q$-periodic  set of local minima separated
by ever-rising barriers.
This might lead to a complex pattern of metastability,
which is a characteristic of the glass phase.

In conclusion, there is the possibility that this Matrix Model might describe 
some of the expected effects 
of frustrated magnetic systems in finite dimension 
$D$  \cite{spinglass,wiegmann}.
Moreover, it should be solvable by known techniques  \cite{matmod}.
The frustrated $XY$ model is another system which might develop
these effects in the low-temperature phase; it would be interesting to
pursue the relation between the $XY$ model and the Matrix Model beyond
the planar limit.

\newpage

{\large\bf Acknowledgements}

We would like to thank Philippe Di Francesco, Enzo Marinari,
Giorgio Parisi and Paul Wiegmann for useful discussions.
This work is supported in part by the European Community Program
``Training and Mobility of Researchers'' FMRX-CT96-0012.

\end{document}